\documentclass[twocolumn,showpacs,amsmath,amssymb,nofootinbib]{revtex4}

\usepackage{graphicx}

\begin{document}
\title{Dilemma that cannot be resolved by biased quantum coin flipping}
\author{Satoshi Ishizaka}
\affiliation{Nano Electronics Research Laboratories, NEC Corporation,
34 Miyukigaoka, Tsukuba, 305-8501, Japan}
\date{\today}
%

\begin{abstract}
We show that a biased quantum coin flip (QCF) cannot provide the performance of
a black-boxed biased coin flip, if it satisfies some fidelity conditions.
Although such a QCF satisfies the security conditions of a biased coin flip, it
does not realize the ideal functionality, and therefore, does not fulfill the
demands for universally composable security. Moreover, through a
comparison within a small restricted bias range, we show that an arbitrary QCF
is distinguishable from a black-boxed coin flip unless it is unbiased on both
sides of parties against insensitive cheating. We also point out the difficulty
in developing cheat-sensitive quantum bit commitment in terms of the
uncomposability of a QCF.
\end{abstract}

%
\maketitle
%

Consider Alice and Bob who have just divorced. They agree to flip a coin to
decide who gets their car, but they live in different cities. How do they
flip a coin by telephone? This is a well-known introduction to a coin flip
\cite{Blum82a}, which is now an important cryptographic primitive on
a communication network. Another important primitive is bit commitment (BC).
The purpose of BC is to realize the scenario in which Alice commits to a bit
($b$) and later she reveals it; this is done such that Bob cannot know $b$
until Alice reveals it and she must reveal $b$ as it is. A fair coin flip
is realized by using secure BC in the following way: Alice first commits to
$b$, Bob next sends $b'$ to Alice, and she reveals $b$. Then, $b\!\oplus\!b'$
is a fair random bit, because Bob cannot know $b$ before choosing $b'$ and
Alice cannot change $b$ after knowing $b'$.

%

An effort was made to construct unconditionally secure quantum BC (QBC), but
unfortunately it was shown that all previously proposed QBC protocols are
broken by the so called entanglement attack \cite{Lo97bMayers97a}. After
controversial discussions, it was then generally accepted that 
unconditionally secure QBC is impossible \cite{Dariano06a}. It was also
proved that a perfectly fair quantum coin flip (QCF) is impossible
\cite{Lo98a,Mayers99a,Ambainis04a,Ambainis04b}.

%

Fortunately, quantum mechanics enables biased coin flipping
\cite{Aharonov00a,Spekkens02a,Ambainis04a,Mochon04a,Colbeck07a}.
In a biased QCF, if both Alice and Bob are honest, the outcome is either 0 or
1, each with probability 1/2. A dishonest party can cheat to bias the
probability to $1/2\!+\!\epsilon$, but it is ensured that the amount of the
bias satisfies $|\epsilon|\!<\!1/2$; so a dishonest party cannot fully
control the outcome. Moreover, when a dishonest party tries to largely bias the
probability, a honest party sometimes obtains the outcome {\em reject}, which
conclusively identifies the presence of cheating. In this paper, however, we
only consider {\em insensitive cheating} such that the outcome
{\em reject} never occurs. Even through insensitive cheating, a dishonest party
can generally bias the probability, whose maximum (or minimum) is called the
threshold for cheat sensitivity \cite{Spekkens02a}.

%

On the other hand, let us imagine ideal biased coin flipping such that a
black-box outputs a common random bit to both parties. A dishonest party can
bias the probability of the outcome but can do nothing else because the party
cannot touch the inside of the box at all. A biased QCF, at first glance,
realizes the black-boxed coin flip, because it is ensured by the laws of
physics that the bias range is limited to $|\epsilon|\!<\!1/2$ against all
possible operations for cheating.

%

In this paper, however, we show that a biased QCF generally does not provide
the performance of a black-boxed biased coin flip, when it is used to
resolve a quantum dilemma. Although such a QCF satisfies the security
conditions of a biased coin flip, it does not realize the ideal functionality.
This warns that, if a QCF is combined with another quantum cryptographic
protocol, an unexpected security hole will occur.

%

Now, let us introduce a quantum dilemma where, in some sense, the car in the
previous dilemma concerning divorce is replaced with a fully quantum object:
an entangled state. Suppose that Alice is required to send half of a maximally
entangled state to Bob. However, Bob is doubtful whether she sends the
entangled state honestly. On the other hand, dishonest Bob sometimes destroys
the shared entanglement and Alice worries about this. At a later time, Bob
wishes to confirm that Alice has honestly sent the
entangled state, and Alice wishes to confirm that the entanglement has been
maintained safely. Therefore, both wish to get the whole state in her/his hand,
because the entanglement cannot be evaluated from their half of the state (this
situation is analogous to quantum bit escrow \cite{Aharonov00a} as we will
discuss later).

%

Since both wishes cannot be satisfied simultaneously, let us introduce a coin
flip to resolve this dilemma, and thus consider the following protocol:

%

{\it Protocol 1 (sharing and maintaining entanglement)}

{\it Stage 1 (sharing):} Alice prepares
$|\phi\rangle_{AB}\!=\!(|00\rangle\!+\!|11\rangle)/\sqrt{2}$ and sends the
$B$ qubit to Bob. This maximal entanglement is to be shared and maintained.

{\it Stage 2 (coin flip):} Alice and Bob execute a coin flipping subprotocol.
If the output of the subprotocol is 0 (1), Alice (Bob) loses the coin flip.

{\it Stage 3 (verification):}  The winner of the coin flip obtains both $A$ and
$B$ qubits and checks whether or not the state of the $AB$ qubits is
$|\phi\rangle$ by a projective measurement. If the state is not
$|\phi\rangle$, the party detects the cheating of the other party with
nonzero probability.

%

Suppose that Alice is dishonest and sends a partially entangled state
$|\Phi(a)\rangle_{AB}\!=\!\sqrt{a}|00\rangle\!+\!\sqrt{1\!-\!a}|11\rangle$
($1/2\!<\!a\!\le\!1$) instead of $|\phi\rangle$ in the stage 1. Let $P_d$
be the probability that Bob detects this cheating. The performance of the
protocol is characterized by the minimal value of $P_d$ for a given $a$. Let us
consider the case where a black-boxed coin flip is used in
the stage 2. The allowable maximal (minimal) bias of the probability of the
outcome 0 is $\epsilon_{\rm max}$ ($\epsilon_{\rm min}$) and
$\epsilon_{\rm min}\!<\!0\!<\!\epsilon_{\rm max}$. Exploiting this
controllable bias range, Alice tries to decrease $P_d$. However, as proved
later, the best strategy is to constantly bias the probability to
$1/2\!+\!\epsilon_{\rm min}$, and to send the $A$ qubit as it is in the
stage 3, when she loses the coin flip. Namely,
\begin{eqnarray}
P_d \ge P^{\rm box}_d&\!\equiv\!&
\textstyle(\frac{1}{2}\!+\!\epsilon_{\rm min})
\hbox{tr}\big[
|\Phi(a)\rangle\langle\Phi(a)|(I\!-\!|\phi\rangle\langle\phi|)\big] \cr
&\!=\!&
\textstyle(\frac{1}{2}\!+\!\epsilon_{\rm min})(\frac{1}{2}-\sqrt{a(1-a)}).
\label{eq: Classical Pd}
\end{eqnarray}

%

Our concern is whether or not a QCF can provide the same performance. To
investigate this, let us recall a unitary model of a QCF
\cite{Ambainis04a,Ambainis04b}, where all classical communication is replaced
by quantum communication and all measurements are postponed until the end of
the protocol. We can thus assume that Alice or Bob's operation in each round
is a
unitary transformation. Following the model in \cite{Ambainis04a}, let
$|\psi_{\rm ini}\rangle$ be an initial state of the protocol. Alice first
applies $U_1$ to her own qubits and sends some qubits to Bob, and then Bob
applies $U_2$ and sends some qubits to Alice. They repeat this and the final
state
after all the rounds is
$|\psi_{\rm fin}\rangle\!=\!(\cdots U_3U_2U_1)|\psi_{\rm ini}\rangle$.
Alice and Bob then measure $|\psi_{\rm fin}\rangle$ to obtain the outcome. When
both are honest, they can obtain 0 or 1 with probability 1/2, and so
$|\psi_{\rm fin}\rangle$ is decomposed such that
$|\psi_{\rm fin}\rangle\!=\!|\psi_0\rangle\!+\!|\psi_1\rangle$, where
$|\psi_c\rangle$ is a part of leading to the outcome $c$,
$\langle\psi_0|\psi_1\rangle\!=\!0$, and $\left\|\psi_c\right\|^2\!=\!1/2$.
Moreover, since both must know the outcome certainty,
$F(\varrho_{X,0},\varrho_{X,1})\!=\!0$, where
$F(\varrho,\sigma)\!=\!(\hbox{tr}\sqrt{\varrho^{1/2}\sigma\varrho^{1/2}})^2$
is the fidelity and $\varrho_{X,c}$ is the normalized reduced state of
$|\psi_c\rangle$ for the party $X\!=\!A,B$ \cite{Ambainis04a}.

%

Let $\epsilon_{\rm max}$ be the possible maximal bias for the outcome 0 of
this QCF, which is achieved if Alice applies $U'_i$ instead of $U_i$. The final
state is then
$|\psi'_{\rm fin}\rangle\!=\!(\cdots U'_3U_2U'_1)|\psi_{\rm ini}\rangle
\!=\!|\psi'_0\rangle\!+\!|\psi'_1\rangle$,
where $\langle\psi'_0|\psi'_1\rangle\!=\!0$ and
$\left\|\psi'_0\right\|^2\!=\!1/2\!+\!\epsilon_{\rm max}$. Moreover,
$\varrho'_{B,c}$, which is the reduced state of $|\psi'_c\rangle$, must not be
conclusively distinguished from $\varrho_{B,c}$ by Bob so that the cheating is
insensitive [hence
$\hbox{supp}(\varrho'_{B,c})\!\subset\!\hbox{supp}(\varrho_{B,c})$ where
$\hbox{supp}(\varrho)$ denotes the support space of $\varrho$]. Since Alice
must know the outcome certainty, $F(\varrho'_{A,0},\varrho'_{A,1})\!=\!0$;
otherwise the cheating is sensitive due to the disagreement of their outcomes.
Likewise, let $\epsilon_{\rm min}$ be the minimal bias that is achieved by
$U''_i$. The corresponding final state, the reduced state, and so on, are also
indicated by the double prime. These satisfy the same conditions as in the
$\epsilon_{\rm max}$ case, except
$\left\|\psi''_0\right\|^2\!=\!1/2\!+\!\epsilon_{\rm min}$.

%

Now, let us consider Alice's cheating strategy for the protocol 1. In the
QCF subprotocol executed in the stage 2, she applies the controlled unitary
transformations 
\begin{equation}
O_i=|0\rangle\langle0|_A\!\otimes\!U''_i
+|1\rangle\langle1|_A\!\otimes\!U'_i
\label{eq: Cheating}
\end{equation}
to $|\Phi(a)\rangle_{AB}\!\otimes\!|\psi_{\rm ini}\rangle$. The whole state
after all the rounds of the QCF subprotocol is
\begin{eqnarray}
\lefteqn{\hspace{-0.8cm}\sqrt{a}|00\rangle_{AB}\!\otimes\!(|\psi''_0\rangle\!+\!|\psi''_1\rangle)
+\sqrt{1\!-\!a}|11\rangle_{AB}\!\otimes\!(|\psi'_0\rangle\!+\!|\psi'_1\rangle)}  \cr
&\!=\!&\sqrt{a}|00\rangle_{AB}\!\otimes\!|\psi''_0\rangle
+\sqrt{1-a}|11\rangle_{AB}\!\otimes\!|\psi'_0\rangle \cr
&\!+\!&\sqrt{a}|00\rangle_{AB}\!\otimes\!|\psi''_1\rangle
+\sqrt{1-a}|11\rangle_{AB}\!\otimes\!|\psi'_1\rangle.
\label{eq: final state}
\end{eqnarray}
The first two and the last two terms in Eq.\ (\ref{eq: final state}) lead to
the outcomes 0 and 1, respectively. Since Bob's reduced state of the system
employed for the QCF is $a\varrho''_{B,c}\!+\!(1\!-\!a)\varrho'_{B,c}$ for the
outcome $c$, and
$\hbox{supp}(\varrho''_{B,c}),\hbox{supp}(\varrho'_{B,c})\!
\subset\!\hbox{supp}(\varrho_{B,c})$,
he knows the outcome certainty by a regular measurement. Alice's reduced
state is
$a|0\rangle\langle0|_A\!\otimes\!\varrho''_{A,c}\!+\!
(1\!-\!a)|1\rangle\langle1|_A\!\otimes\!\varrho'_{A,c}$,
and she can obtain the outcome certainty using the projector
$|0\rangle\langle0|_A\!\otimes\!\Pi''_c\!+\!
|1\rangle\langle1|_A\!\otimes\!\Pi'_c$,
where $\Pi'_c$ ($\Pi''_c$) distinguishes $\varrho'_{A,0}$ and $\varrho'_{A,1}$
($\varrho''_{A,0}$ and $\varrho''_{A,1}$). These projectors exist because
$F(\varrho'_{A,0},\varrho'_{A,1})\!=\!
F(\varrho''_{A,0},\varrho''_{A,1})\!=\!0$ \cite{Ambainis04a}.
Suppose that the outcome of the QCF is 0; the state of the $AB$ qubits will be
checked by Bob in the stage 3. Before Alice sends the $A$ qubit to Bob, she
applies
$|0\rangle\langle0|_A\!\otimes\!I\!+\!|1\rangle\langle1|_A\!\otimes\!V$, where
$V$ maximizes the overlap between $|\psi''_0\rangle$ and $|\psi'_0\rangle$
such that
$|\langle\psi''_0|V|\psi'_0\rangle|^2\!=\!\left\|\psi''_0\right\|^2
\left\|\psi'_0\right\|^2F(\varrho'_{B,0},\varrho''_{B,0})$ \cite{Jozsa94a}.
Through this procedure, the whole state becomes
$|\Psi_0\rangle\!=\!\sqrt{a}|00\rangle_{AB}\otimes|\psi''_0\rangle\!+\!
\sqrt{1\!-\!a}|11\rangle_{AB}\otimes V|\psi'_0\rangle$, and $P_d$ in this
strategy is
\begin{eqnarray}
P^{Q}_d&\!=\!&\hbox{tr}\big[|\Psi_0\rangle\langle\Psi_0|
(I-|\phi\rangle\langle\phi|)_{AB}\big] \cr
&\!=\!&\textstyle
\frac{1}{2}\big[
a(\frac{1}{2}\!+\!\epsilon_{\rm min})
+(1\!-\!a)(\frac{1}{2}\!+\!\epsilon_{\rm max})\big] \cr
&&-\big[\textstyle a(1\!-\!a)(\frac{1}{2}\!+\!\epsilon_{\rm min})
(\frac{1}{2}\!+\!\epsilon_{\rm max})F\big]^{1/2}
\label{eq: Quantum Pd}
\end{eqnarray}
where $F\!\equiv\!F(\varrho'_{B,0},\varrho''_{B,0})$. Comparing
Eqs.\ (\ref{eq: Classical Pd}) and (\ref{eq: Quantum Pd}), it is found that
$P^Q_d\!<\!P^{\rm box}_d$ if
$1\!>\!a\!>\!\frac{(r\!-\!1)^2}{4(\sqrt{rF}\!-\!1)^2\!+\!(r-1)^2}$ and
\begin{equation}
F>1/r\equiv(1+2\epsilon_{\rm min})/(1+2\epsilon_{\rm max}).
\label{eq: Threshold of F}
\end{equation}
This result shows that, if a QCF has the property of
Eq.\ (\ref{eq: Threshold of F}), there exists a finite range of $a$ in which
$P^Q_d\!<\!P^{\rm box}_d$. Therefore, it is concluded that such a QCF cannot
provide the performance of a black-boxed coin flip.

%

The point of the above cheating strategy is that it is possible to superpose
two biasing operations $U'_i$ and $U''_i$. This enables to correlate the state
of the $AB$ qubits with the outcome of the QCF such that the state is more
entangled than $|\Phi(a)\rangle_{AB}$ whenever Alice loses the QCF (and thus
$P_d$ decreases). For this purpose, Alice utilizes the difference of
$\left\|\psi''_0\right\|$ and $\left\|\psi'_0\right\|$ ({\it i.e.,} difference
of $\epsilon_{\rm max}$ and $\epsilon_{\rm min}$). However, this
procedure has created undesired entanglement between the $AB$ qubits and the
system employed for the QCF, and so Alice needs to disentangle them; otherwise
the entanglement of the $AB$ qubits will be washed out by the undesired
entanglement. This is done by increasing the overlap between $|\psi'_0\rangle$
and $|\psi''_0\rangle$. Note that the disentangling process is incomplete
(unless $F\!=\!1$), and as a result, the state of the $AB$ qubits (the reduced
state of $|\Psi_0\rangle$) is a mixed state.
Equation (\ref{eq: Threshold of F}) agrees
with the condition that this mixed state is more entangled than
$|\Phi(a)\rangle$ in the measure of negativity \cite{Vidal02a}.

%

As shown above, a QCF does not provide the performance of a black-boxed coin
flip, if it satisfies Eq.\ (\ref{eq: Threshold of F}). To further investigate
this, let us introduce the following biasing operation: Suppose that
Alice has a local ancilla qubit and she prepares the initial state 
$|\psi_{\rm ini}\rangle\!\otimes\!
(\sqrt{1\!-\!x}|0\rangle_a\!+\!\sqrt{x}|1\rangle_a)$
for the QCF, where the subscript $a$ denotes the ancilla qubit. Then, if she
applies
$\tilde U_i\!=\!U_i\!\otimes\!|0\rangle\langle0|_a
\!+\!U''_i\!\otimes\!|1\rangle\langle1|_a$
to the initial state [do not confuse this with Eq.\ (\ref{eq: Cheating})],
the QCF
is biased by $x\epsilon_{\rm min}$. Moreover, when the outcome is 0, Bob's
reduced state is
$\tilde\varrho_{B,0}
\!=\!\varrho_{B,0}\!+\!x(\varrho''_{B,0}\!-\!\varrho_{B,0})$,
and hence
$F(x)\!\equiv\!F(\varrho_{B,0},\tilde\varrho_{B,0})
\!=\!1\!-\!{\cal O}(x^2)$ \cite{Note_fidelity}.
Now, let us imagine a special circumstance where Alice's ability is restricted
such that she can only use $\tilde U_i$ for biasing the QCF (the point is that
she cannot directly employ $U''_i$). As a result, the bias of the QCF is
restricted within $[x\epsilon_{\rm min},0]$, and therefore, it may be
natural to compare it with the black-boxed coin flip with the same bias
range. Then, if Alice adopts the cheating strategy like
Eq.\ (\ref{eq: Cheating}), where $U_i$ and $\tilde U_i$ are superposed as
$O_i\!=\!|0\rangle\langle0|_A\!\otimes\!\tilde U_i
\!+\!|1\rangle\langle1|_A\!\otimes\!U_i$ to decrease $P_d$,
we have Eq.\ (\ref{eq: Threshold of F}) in which $\epsilon_{\rm max}$ and
$\epsilon_{\rm min}$ is replaced by 0 and $x\epsilon_{\rm min}$,
respectively; so $P^{Q}_d\!<\!P^{\rm box}_d$ if
$F(x)\!>\!1\!+\!2x\epsilon_{\rm min}$. However, this fidelity condition is
always satisfied for
$x\!\rightarrow\!0$ because $F(x)\!=\!1\!-\!{\cal O}(x^2)$
as mentioned above. The same discussion holds if the bias is restricted within
$[0,x\epsilon_{\rm max}]$. In this way, an arbitrary QCF is
distinguishable from a black-boxed coin flip (as $P^Q_d\!<\!P^{\rm box}_d$)
unless the QCF is unbiased against insensitive cheating
(if we compare them around $\epsilon\!=\!0$).

%

\begin{figure}[t]
\centerline{\scalebox{0.5}[0.5]{\includegraphics{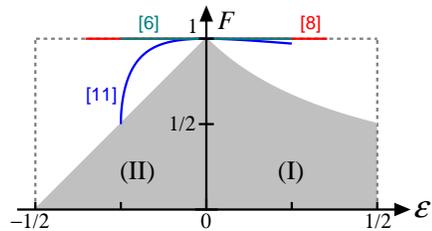}}}
\caption{Bound for fidelity ($F$) as function of bias ($\epsilon$). If a QCF
is located outside the gray region, it is distinguishable from a black-boxed
coin flip. The fidelity for the QCF protocols proposed in
\protect\cite{Aharonov00a}, \protect\cite{Ambainis04a}, and
\protect\cite{Colbeck07a} is also plotted, where we assumed dishonest Bob. Two
ends of each line correspond to the maximal and minimal bias for each QCF
protocol.
}
\label{fig: 1}
\end{figure}

To see these results graphically, the following two bounds are plotted in
Fig.\ \ref{fig: 1}:
$$
\left.
\begin{array}{cll}
\hbox{(I)}& F(\epsilon)>1/(1+2\epsilon)&\hbox{for $\epsilon_{\rm min}\!=\!0$ and
$\epsilon\!=\!x\epsilon_{\rm max}\!\ge\!0$,} \cr
\hbox{(II)} & F(\epsilon)>1+2\epsilon&\hbox{for $\epsilon_{\rm max}\!=\!0$ and
$\epsilon\!=\!x\epsilon_{\rm min}\!\le\!0$.}
\end{array}
\right.
$$
If the fidelity $F$ of the QCF, whose bias is forcedly restricted within
(I) $[0,\epsilon]$ and (II) $[\epsilon,0]$, is located outside the gray
region, the QCF is distinguishable from the black-boxed coin flip with the
same bias range. The fidelity for
some of the proposed QCF protocols is also plotted for a comparison.

%

All of the above discussions hold when Bob is dishonest. This is because the
protocol 1 is essentially symmetric with respect to parties, if we assume
that Bob's dishonest action in the stage 1 is to perform the following positive
operator valued measurement (POVM) of the $B$ qubit:
\begin{eqnarray}
M_0&\!=\!&\sqrt{a}|0\rangle\langle0|+\sqrt{1\!-\!a}|1\rangle\langle1|, \cr
M_1&\!=\!&\sqrt{1\!-\!a}|0\rangle\langle0|+\sqrt{a}|1\rangle\langle1|,
\label{eq: POVM}
\end{eqnarray}
where $M_0^\dagger M_0\!+\!M_1^\dagger M_1\!=\!\openone_B$. Depending on the
outcome of the POVM, the post-measured state becomes $|\Phi(a)\rangle_{AB}$
or $|\Phi(1\!-\!a)\rangle_{AB}$, each with probability 1/2. He then tries to
decrease $P_d$. For $|\Phi(a)\rangle$, the same cheating strategy as used
with dishonest Alice is applicable. This is the case for
$|\Phi(1\!-\!a)\rangle$, if the role
of $|0\rangle_B$ and $|1\rangle_B$ is exchanged in the controlled operations of
the cheating strategy. Then, we have the same bound for
$F\!\equiv\!F(\varrho'_{A,1},\varrho''_{A,1})$, but $\epsilon_{\rm max}$ and
$\epsilon_{\rm min}$ must be read as those for the outcome 1 of the QCF.
This implies that a QCF must be unbiased on both Alice and Bob's sides
simultaneously, so that it is indistinguishable from a black-boxed coin flip.

%

So, let us now prove Eq.\ (\ref{eq: Classical Pd}). The general action of
dishonest Alice when deciding on the bias of a black-boxed coin flip is
described by a POVM $\{L_\epsilon\}$ of the $A$ qubit
($\int\!d\epsilon L^{\dagger}_{\epsilon}L_{\epsilon}\!=\!\openone_A$).
The probability of the outcome 0 is then biased to
$\frac{1}{2}\!+\!\epsilon$, and $|\Phi(a)\rangle_{AB}$ will be checked by
Bob with this probability. Before sending the $A$ qubit, she can apply a
trace-preserving operation regarding $\epsilon$, but this is included in
$L_\epsilon$. Moreover, the singlet fraction
${\cal F}(\sigma)\!=\!\langle\phi|\sigma|\phi\rangle$ is bounded as
${\cal F}(\sigma)\!\le\![\hbox{tr}\sigma\!+\!N_B(\sigma)]/2$, where
$N_B(\sigma)$ is negativity \cite{Vidal02a} [the subscript denotes the
partial transposition with respect to the $B$ qubit].
Since $N_B$ is an entanglement monotone \cite{Plenio05a}, the average cannot
be increased by the local operation of the POVM. Hence,
\begin{eqnarray}
P_d
&\!\!=\!\!&\!\int_{\epsilon_{\rm min}}^{\epsilon_{\rm max}}
\hspace{-0.6cm}d\epsilon(\textstyle\frac{1}{2}\!+\!\epsilon)
\hbox{tr}\big[
L_\epsilon|\Phi(a)\rangle\langle\Phi(a)|L^{\dagger}_\epsilon
(I\!-\!|\phi\rangle\langle\phi|)\big] \cr
&\!\!\ge\!\!&\textstyle\frac{1}{2}(\frac{1}{2}\!+\!\epsilon_{\rm min})
\big[1\!-\!\displaystyle\int\!d\epsilon
N_B(
L_\epsilon|\Phi(a)\rangle\langle\Phi(a)|L^{\dagger}_\epsilon
)\big] \cr
&\!\!\ge\!\!&\textstyle\frac{1}{2}(\frac{1}{2}\!+\!\epsilon_{\rm min})
\big[1\!-\!N_B(|\Phi(a)\rangle)\big],
\label{eq: Deriving Pd}
\end{eqnarray}
and we have Eq.\ (\ref{eq: Classical Pd}), because
$N_B(|\Phi(a)\rangle)\!=\!2\sqrt{a(1\!-\!a)}$.

%

So far, we have focused on the comparison through $P_d$. Now, we
concentrate on a case where $P_d\!=\!0$; the probability of
detecting cheating is strictly zero, and so the state of the $AB$ qubits must
be precisely $|\phi\rangle$ when it is checked in the stage 3. The
performance of the protocol 1 is then characterized by the maximal allowed
value of $a$ for a dishonest party. Suppose again that Alice is dishonest. For
a black-boxed coin flip, it is found from Eq.\ (\ref{eq: Classical Pd}) that
$a\!=\!\frac{1}{2}$ must hold regardless of the bias range; so she cannot cheat
at all under $P_d\!=\!0$, as expected. For a QCF, however, it is found from
Eq.\ (\ref{eq: Quantum Pd}) that $P^{Q}_d\!=\!0$ even for $a\!>\!1/2$ if
$F\!=\!1$. This occurs for an arbitrary pair of biasing operations as far as
the pair of operations satisfies $F\!=\!1$. So, by replacing
$\epsilon_{\rm max}$ and $\epsilon_{\rm min}$ in
Eq.\ (\ref{eq: Quantum Pd}) with $\epsilon'$ and $\epsilon''$,
respectively, we have $P^{Q}_d\!=\!0$ if 
$a\!=\!1/2\!+\!(\epsilon'\!-\!\epsilon'')
/[2(1\!+\!\epsilon'\!+\!\epsilon'')]$
and $F(\varrho'_{B,0},\varrho''_{B,0})\!=\!1$. Basically, if the QCF has a
property of
\begin{equation}
\Delta a\equiv\max_{\epsilon',\epsilon'',F=1}
\frac{\epsilon'\!-\!\epsilon''}
{2(1\!+\!\epsilon'\!+\!\epsilon'')}>0,
\label{eq: Delta a}
\end{equation}
Alice can successfully cheat because $a\!=\!1/2\!+\!\Delta a\!>\!1/2$ while $P^{Q}_d\!=\!0$.
The maximization in Eq.\ (\ref{eq: Delta a}) is taken over all the pairs of 
the two biasing operations
($\epsilon'$ and $\epsilon''$) subject to
$F(\varrho'_{B,0},\varrho''_{B,0})\!=\!1$.
Such a QCF also cannot provide the performance of a black-boxed coin flip, and
even allows the cheating that is completely prohibited by a black-boxed coin
flip. Note that the same discussion holds again for dishonest Bob
[$\epsilon'$ and $\epsilon''$ are those for
the outcome 1 and $F(\varrho'_{A,1},\varrho''_{A,1})\!=\!1$].

%

As a simple example, let us analyze the following protocol \cite{Hardy04a}
(this is not a true QCF because the probability of the outcome is not 1/2
even if both are honest, but the above cheating strategy is applicable):

%

{\it Protocol 2 (QCF like):}
Alice prepares $|\phi\rangle_{CD}$ and sends the $D$ qubit to Bob. He
optionally checks $|\phi\rangle_{CD}$ (getting the $C$ qubit). If he uses the
option, this protocol automatically outputs 1. Otherwise, he measures
the $D$ qubit in the $\{|0\rangle,|1\rangle\}$ basis, sends the result to
Alice, and she confirms the validity by measuring the $C$ qubit. This protocol
then outputs the measurement result.

%

In this protocol, it is confirmed that
$F(\varrho'_{A,1},\varrho''_{A,1})\!=\!1$ for Bob's two biasing operations of
(i) he always uses the option ($\epsilon'\!=\!1/2$), and (ii) he measures
the $D$ qubit, and if the result is 1, he uses the option
($\epsilon''\!=\!0$). Hence, we have $\Delta a\!\ge\!1/6$ and $a\!\ge\!2/3$
from Eq.\ (\ref{eq: Delta a}). On the other hand, it can be shown that
$a\!\le\!2/3$ for Bob's general action \cite{Preprint}. Therefore, it is found
that the cheating strategy considered in this paper has optimally maximized
$a$ under $P_d\!=\!0$. This is the case for the 3-round protocol in
\cite{Aharonov00a} ($a\!=\!\cos\frac{\pi}{8}$) and for the optimal
3-round protocol in \cite{Ambainis04a} ($a\!=\!3/4$), for which we assumed
dishonest Bob. Apart from the optimality of the strategy, the QCF protocols of
\cite{Spekkens02a,Mochon04a,Colbeck07a} also have the property of 
$\Delta a\!>\!0$, at least, on either side of parties.

%

As mentioned before, the situation considered in this paper is analogous to
quantum bit escrow \cite{Aharonov00a} (it is in fact regarded as its
entanglement version).

%

{\it Protocol 3 (quantum bit escrow)}

{\it Stage 1 (commitment):} To commit to $b\!=\!0$ ($1$), Alice prepares either
$|0\rangle_B$ or $|-\rangle_B$ ($|1\rangle_B$ or $|+\rangle_B$), each
with probability 1/2, which is written as $|\xi_{bx}\rangle$ where $x$
denotes the encoding basis. She then sends the $B$ qubit to Bob.

{\it Stage 2 (opening):} Alice reveals $b$.

{\it Stage 3 (verification):} Either Alice or Bob obtains the $B$ qubit and
checks whether or not it is $|\xi_{bx}\rangle$ to detect cheating (Alice
reveals $x$ if Bob checks the state).

%

This is a weak variant of QBC such that either Alice or Bob can detect cheating
with nonzero probability. The question of whether or not it is possible to
use a biased QCF for the purpose of deciding which party
will check the $B$ qubit in the stage 3 was raised in \cite{Aharonov00a}.
If this is so, the resultant protocol
is cheat-sensitive QBC (CSQBC) \cite{Aharonov00a,Hardy04a}, which enables both
to detect cheating, albeit with smaller nonzero probability.

%

However, since the resultant CSQBC has the same structure as in the protocol 1,
it struggles with the difference between a QCF and a black-boxed coin flip. For
example, if $\Delta a\!>\!0$, dishonest Bob can steal partial information about
$b$ before the opening stage by a POVM like in Eq.\ (\ref{eq: POVM}) (whose
$\{|0\rangle,|1\rangle\}$ basis is replaced by an appropriate one to steal the
information \cite{Preprint}). Alice cannot detect his cheating because he can
precisely recover $|\xi_{bx}\rangle$ from a state collapsed by the POVM
whenever he loses the QCF, as he recovers $|\phi\rangle$ from $|\Phi(a)\rangle$
or $|\Phi(1\!-\!a)\rangle$. Likewise, if $\Delta a\!>\!0$,
dishonest Alice can change the probability of revealing $b\!=\!0$ in the
opening stage \cite{Note_reveal}. Therefore, a QCF that is combined with bit
escrow should not satisfy Eq.\ (\ref{eq: Delta a}) on both sides of parties.
Unfortunately, this is not the case in the example of CSQBC suggested
in \cite{Aharonov00a}, and even in \cite{Hardy04a}. We described the cheating
method for those in \cite{Preprint}. Note that, as far as we know, an explicit
protocol for secure CSQBC has not been found yet \cite{Preprint}, contrary to
the widespread belief that CSQBC is possible. 

%

To summarize, we considered the problem of sharing and maintaining entanglement
between distrustful parties, and showed that a QCF cannot provide the
performance of a black-boxed coin flip, if it satisfies the fidelity conditions
of Eqs.\ (\ref{eq: Threshold of F}) or (\ref{eq: Delta a}). Such a QCF
obviously does not fulfill the conditions for universally composable (UC)
security \cite{Note_UC}; the demands for ensuring the security of a
cryptographic primitive regardless of how it is used in applications
\cite{BenOr04a}. This result is quite contrast to quantum key distribution
(QKD), where a QKD protocol is automatically UC
secure if it satisfies the general security conditions \cite{BenOr04b}.
Moreover, through a comparison within a small restricted bias range, we showed
that an arbitrary QCF is distinguishable from a black-boxed coin flip unless it
is unbiased on both sides of parties against insensitive cheating, {\it i.e.,}
unless it is a cheat-sensitive unbiased QCF. Finally, we discussed the
relation to CSQBC constructed from bit escrow and a QCF, and pointed out the
difficulty in developing secure CSQBC in terms of the uncomposability condition
of Eq.\ (\ref{eq: Delta a}). We wish these results could shed some light on the
important open problem of whether or not quantum mechanics enables
cheat-sensitive bit commitment and cheat-sensitive unbiased coin flipping.


\begin{thebibliography}{10}
\bibitem{Note_fidelity}
$F(\varrho,\varrho\!+\!x\Delta)\!=\!1\!+\!x\hbox{tr}(\Pi\Delta)\!+\!
{\cal O}(x^2)$ where $\Pi$ is the projector to $\hbox{supp}(\varrho)$.
\bibitem{Note_reveal}
Alice sends the $B$ qubit of
$(|00'\rangle\!+\!|11'\rangle)_{AB}/\sqrt{2}$ in the commitment stage, where
$|0'\rangle\!=\!\alpha|0\rangle\!-\!\beta|1\rangle$,
$|1'\rangle\!=\!\beta|0\rangle\!+\!\alpha|1\rangle$,
$\alpha\!=\!\cos\frac{\pi}{8}$, and $\beta\!=\!\sin\frac{\pi}{8}$.
In the opening stage, if she performs the POVM of
$E_0\!=\!\alpha|0\rangle\langle0|_A\!+\!\beta|1\rangle\langle1|_A$ for
deciding $b$, she can reveal $b\!=\!0$ with
probability 1/2, because the post measured states are
$|\eta_0\rangle\!=\!\alpha|00'\rangle\!+\!\beta|11'\rangle
\!\propto\!|+0\rangle\!+\!|--\rangle$
and $|\eta_1\rangle\!\propto\!|++\rangle\!-\!|-1\rangle$ for $E_0$ and
$E_1\!=\!(I\!-\!E^{\dagger}_0E_0)^{1/2}$,
respectively. However, if she performs the POVM of
$E'_0\!=\!\sqrt{\alpha^2\!+\!\delta}|0\rangle\langle0|_A\!
+\!\sqrt{\beta^2\!+\!\delta}|1\rangle\langle1|_A$, she can reveal $b\!=\!0$
with probability $1/2\!+\!\delta$, because she can precisely change
the post measured state into $|\eta_b\rangle$ whenever she loses the QCF.
The nonzero $\delta$ exists if $\Delta a\!>\!0$.
\bibitem{Note_UC}
In the comparison through $P_d$, the probability of the outcome is also
different from each other. If the probability is adjusted to a QCF, then
$P^{\rm box}_d$ increases, and a black-boxed coin flip cannot simulate the QCF
anyway.
\end{thebibliography}

%
\end{document}